\documentclass[twocolumn]{aastex62}

\usepackage{newtxtext,newtxmath}
\usepackage[T1]{fontenc}
\usepackage{ae,aecompl}
\usepackage{graphicx}	
\usepackage{amsmath}	
\usepackage{amssymb}	
\usepackage{media9}

\hypersetup{
  colorlinks=true,
  urlcolor=blue,
  citecolor= blue,
  linkcolor = blue
}

\graphicspath{{./}{figures/}}

\submitjournal{ApJ}

\shorttitle{\textsc{M92 Stream}}
\shortauthors{Thomas et al.}

\begin{document}

\title{The hidden past of M92: Detection and characterization of a newly formed 17$\degr$ long stellar stream using the Canada-France Imaging Survey}

\correspondingauthor{Guillaume F. Thomas}
\email{guillaume.thomas.astro@gmail.com}

\author[0000-0002-2468-5521]{Guillaume F. Thomas}
\affiliation{NRC Herzberg Astronomy and Astrophysics, 5071 West Saanich Road, Victoria, BC, V9E 2E7, Canada}

\author[0000-0002-4350-7632]{Jaclyn Jensen}
\affiliation{Department of Physics and Astronomy, University of Victoria, Victoria, BC, V8P 1A1, Canada}
\affiliation{NRC Herzberg Astronomy and Astrophysics, 5071 West Saanich Road, Victoria, BC, V9E 2E7, Canada}

\author[0000-0003-4666-6564]{Alan McConnachie}
\affiliation{NRC Herzberg Astronomy and Astrophysics, 5071 West Saanich Road, Victoria, BC, V9E 2E7, Canada}

\author[0000-0003-1184-8114]{Patrick C\^ot\'e}
\affiliation{NRC Herzberg Astronomy and Astrophysics, 5071 West Saanich Road, Victoria, BC, V9E 2E7, Canada}

\author[ 	
0000-0003-4134-2042]{Kim Venn}
\affiliation{Department of Physics and Astronomy, University of Victoria, Victoria, BC, V8P 1A1, Canada}

\author{Nicolas Longeard}
\affiliation{Laboratoire d'astrophysique, \'Ecole Polytechnique F\'ed\'erale de Lausanne (EPFL), Observatoire, 1290 Versoix, Switzerland}

\author[0000-0002-7667-0081]{Raymond Carlberg}
\affiliation{Department of Astronomy and Astrophysics, University of Toronto, 50 St. George Street, Toronto, ON M5S 3H4, Canada}

\author{Scott Chapman}
\affiliation{Department of Physics and Atmospheric Science, Dalhousie University, Halifax, NS, B3H 4R2, Canada}
\affiliation{National Research Council, Herzberg Astronomy and Astrophysics, 5071West Saanich Road, Victoria, V9E 2E7, Canada}
\affiliation{Department of Physics and Astronomy, University of British Columbia, Vancouver, BC, V6T 1Z1, Canada}

\author{Jean-Charles Cuillandre}
\affiliation{AIM, CEA, CNRS, Universit\'e Paris-Saclay, Universit\'e Paris Diderot, Sorbonne Paris Cit\'e, Observatoire de Paris, PSL University, F-91191 Gif-sur-Yvette, France}

\author[0000-0003-3180-9825]{Benoit Famaey}
\affiliation{Universit\'e de Strasbourg, CNRS, Observatoire astronomique de Strasbourg, UMR 7550, F-67000 Strasbourg, France}

\author[0000-0002-8224-1128]{Laura Ferrarese}
\affiliation{NRC Herzberg Astronomy and Astrophysics, 5071 West Saanich Road, Victoria, BC, V9E 2E7, Canada}

\author{Stephen Gwyn}
\affiliation{NRC Herzberg Astronomy and Astrophysics, 5071 West Saanich Road, Victoria, BC, V9E 2E7, Canada}

\author[0000-0002-2165-5044]{Fran\c{c}ois Hammer}
\affiliation{GEPI, Observatoire de Paris, Université PSL, CNRS, Place Jules Janssen F-92195, Meudon, France}

\author[0000-0002-3292-9709]{Rodrigo A. Ibata}
\affiliation{Universit\'e de Strasbourg, CNRS, Observatoire astronomique de Strasbourg, UMR 7550, F-67000 Strasbourg, France}

\author[0000-0002-8318-433X]{Khyati Malhan}
\affiliation{The Oskar Klein Centre, Department of Physics, Stockholm University, AlbaNova, SE-10691 Stockholm, Sweden}

\author[0000-0002-1349-202X]{Nicolas F. Martin}
\affiliation{Universit\'e de Strasbourg, CNRS, Observatoire astronomique de Strasbourg, UMR 7550, F-67000 Strasbourg, France}

\author[0000-0002-2849-559X]{Simona Mei}
\affiliation{Universit\'{e} de Paris, F-75013, Paris, France, LERMA, Observatoire de Paris, PSL Research University, CNRS, Sorbonne Universit\'e,  F-75014 Paris, France}
\affiliation{Jet Propulsion Laboratory, Cahill Center for Astronomy \& Astrophysics, California Institute of Technology, 4800 Oak Grove Drive, Pasadena, CA, USA}

\author{Julio F. Navarro}
\affiliation{Department of Physics and Astronomy, University of Victoria, Victoria, BC, V8P 1A1, Canada}

\author{C\'eline Reyl\'e}
\affiliation{Institut UTINAM, CNRS UMR6213, Univ. Bourgogne Franche-Comt\'e, OSU THETA Franche-Comt\'e-Bourgogne, Observatoire de Besan\c con, BP 1615, 25010 Besan\c con Cedex, France}

\author{Else Starkenburg}
\affiliation{Leibniz Institute for Astrophysics Potsdam (AIP), An der Sternwarte 16, D-14482 Potsdam, Germany}

\begin{abstract}
We present an analysis of the structure, kinematics and orbit of a newly found stellar stream emanating from the globular cluster M92 (NGC 6341). This stream was discovered in an improved matched-filter map of the outer Galaxy, based on a "color-color-magnitude"  diagram, created using photometry from the Canada-France Imaging Survey (CFIS) and the Pan-STARRS 1 3$\pi$ survey (PS1). We find the stream to have a length of $17\degr$ (2.5 kpc at the distance of M92), a width dispersion of 0.29$\degr$(42 pc) and a stellar mass of $[3.17 \pm{0.89}] \times 10^{4} $ M$_\odot$ ($10 \%$ of the stellar mass of the current main body of M92). We examine the kinematics of main sequence, red giant and blue horizontal branch stars belonging to the stream and that have proper motion measurements from the second data release of Gaia. N-body simulations suggest that the stream was likely formed very recently (during the last $\sim 500$ Myr)  forcing us to question the orbital origin of this ancient, metal-poor globular cluster.
\end{abstract}

\keywords{globular clusters: individual: M 92 - Galaxy: kinematics and dynamics - Galaxy: halo - Galaxy: formation}

\section{Introduction}
Thin and dynamically cold stellar streams are formed by the disruption of low-mass progenitors, such as globular clusters, through tidal effects or disk shocking in a host galaxy \citep[e.g.][]{combes_1999,johnston_1999}. These thin structures have proved to be very valuable tracers of the Galactic potential and consequently of the mass distribution of the Milky Way \citep[e.g.][]{dehnen_2004,bonaca_2014,kupper_2015,pearson_2015,thomas_2017,thomas_2018,bonaca_2018,malhan_2019}, while also potentially being direct witnesses of the hierarchical formation of the Galaxy \citep{johnston_2008}. For these reasons, the more stellar streams detected and characterized, the tighter the constraints will be on the three dimensional Galactic potential as a function of radius.

\begin{figure*}
\centering
\includegraphics[angle=0, clip,width=18.0cm]{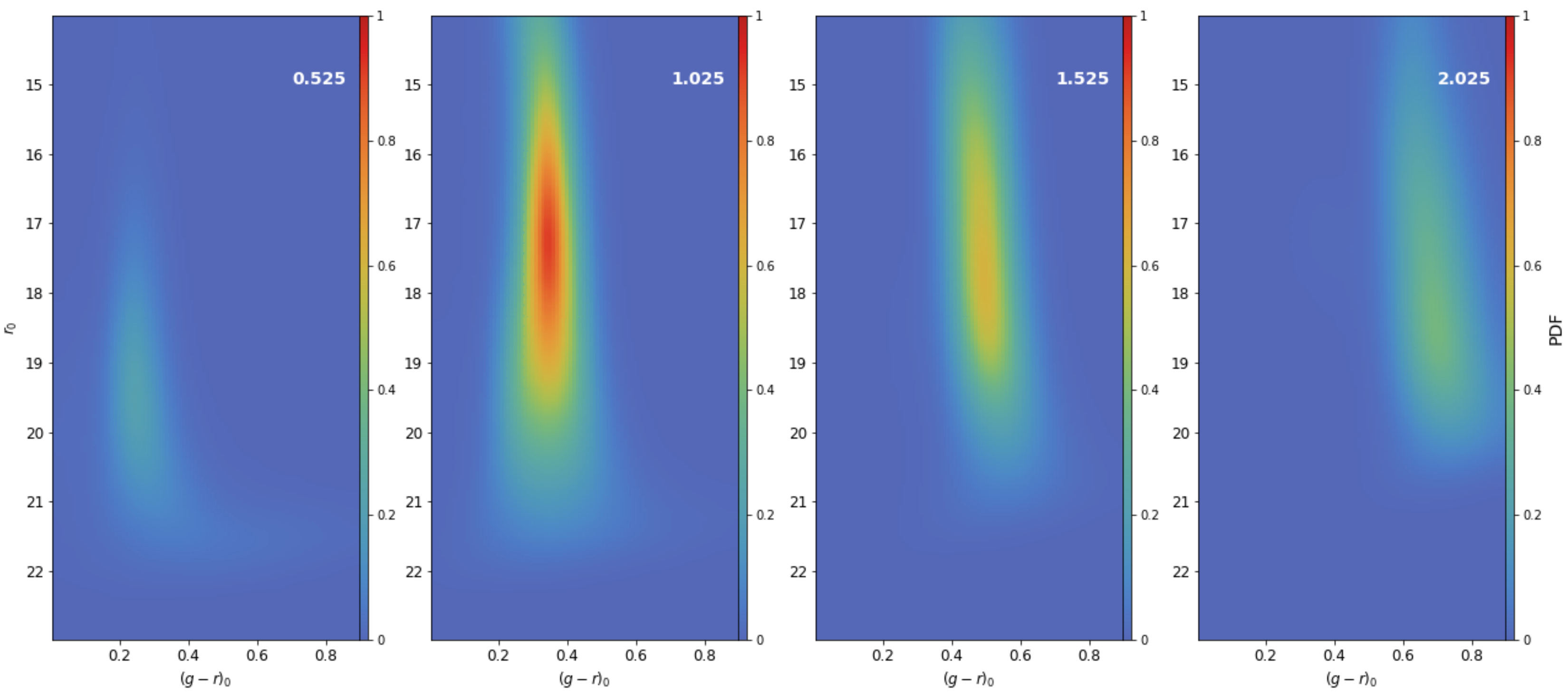}
   \caption{Representation of the color-color-magnitude diagram (CCMD) of the field stars by different color-magnitude-diagram for different value of $(u-g)_0$, whose the value are indicated in the upper right side of each panel.}
\label{CCMD}
\end{figure*}

In addition, globular clusters streams are sensitive to small-scale variations in the Galactic potential, making them promising probes of the granularity of the dark matter halo. This is in contrast to other dynamical tracers, which are often only sensitive to the integrated mass within a given radius \citep[the global kinematics of globular clusters and dwarf galaxies; e.g.][]{deason_2012,eadie_2017,monari_2018}. Indeed, the distribution of stars along these streams can be affected by external perturbations produced by the Galactic bar \citep{hattori_2016,pearson_2017}, spiral arms \citep{banik_2019}, giant molecular clouds \citep{amorisco_2016}, dark matter sub-haloes \citep[e.g.][]{johnston_2002,ibata_2002,carlberg_2012,ngan_2015,erkal_2015a,bonaca_2019} and, more likely, a combination of all of them. It can be difficult to distinguish the signatures of these effects from those produced by the internal dynamics of the cluster itself, such as possible degeneracies between the effects of substructures and those of internal epicyclic motions \citep{kupper_2008,kupper_2010,kupper_2012,mastrobuono-battisti_2012,mastrobuono-battisti_2013,ibata_2020}. Furthermore, it is important to keep in mind that the streams are faint and cover several degrees on the sky, and some of the observed variations in the inner structure of a stream might actually be artificial, consequences of the inhomogeneities of large observational surveys \citep{thomas_2016,ibata_2020}. 

For all of these reasons, it is crucial to have a statistically significant sample of extended globular clusters streams. In the last few years, the number of known streams around the Milky Way has increased drastically \citep[see the review of][]{newberg_2016}, thanks to the advent of large surveys such as Pan-STARRS 3 $\pi$ (PS 1) and the Dark Energy Survey (DES) \citep{balbinot_2016,bernard_2016,grillmair_2017,myeong_2017,navarrete_2017,mateu_2018,shipp_2018}. In addition, a great number of streams have been discovered using new methods exploiting the proper motions of the second Gaia data release \citep{malhan_2018,bianchini_2019,carballo-bello_2019,grillmair_2019,ibata_2019,ibata_2019a,palau_2019,sollima_2020}. At the moment $\sim 40$ globular clusters streams are observed around the Milky Way, with Galactocentric distances ranging from 1 to 45 kpc. However, only a couple of the streams that cover more than a few degrees have an obvious progenitor, in the form of a surviving globular cluster (e.g., Palomar 5 and 15, M5, M68, NGC 5466, NGC 7492 and $\omega$-Centauri; see references above). Knowledge of the progenitor properties is useful in reducing the number of free parameters when modelling these streams. Thus, finding additional streams with unambiguous progenitors will be useful for probing both the shape of the Galactic potential and its granularity.

In this paper we present the detection of a 17$\degr$ long stellar stream around the M92 globular cluster and characterize its properties using a suite of dynamical models. The presence of a stream emanating from M92 was originally predicted by \citet{balbinot_2018}, based on the analysis of the orbital and dynamical properties of the cluster. During the preparation of this manuscript, a part of this stream was independently detected by \citet{sollima_2020}. Section \ref{sec_method} presents the data and the matched-filter method used to detect the stream. Section \ref{sec_results} presents an analysis of the stream and a kinematic confirmation of its existence using other stellar tracers. A suite of dynamical models and simulations of this stream, used to estimate its dynamical age, are described in Section \ref{sec_dynmod} and the results are discussed in Section \ref{sec_discussion}. Finally, we summarize our results and draw our conclusions in Section \ref{sec_conclusion}.

\section{Method} \label{sec_method}
\subsection{The data}

\begin{figure*}
\centering
  \includegraphics[angle=0,viewport= 0 0 1025 455, clip,width=18.0cm]{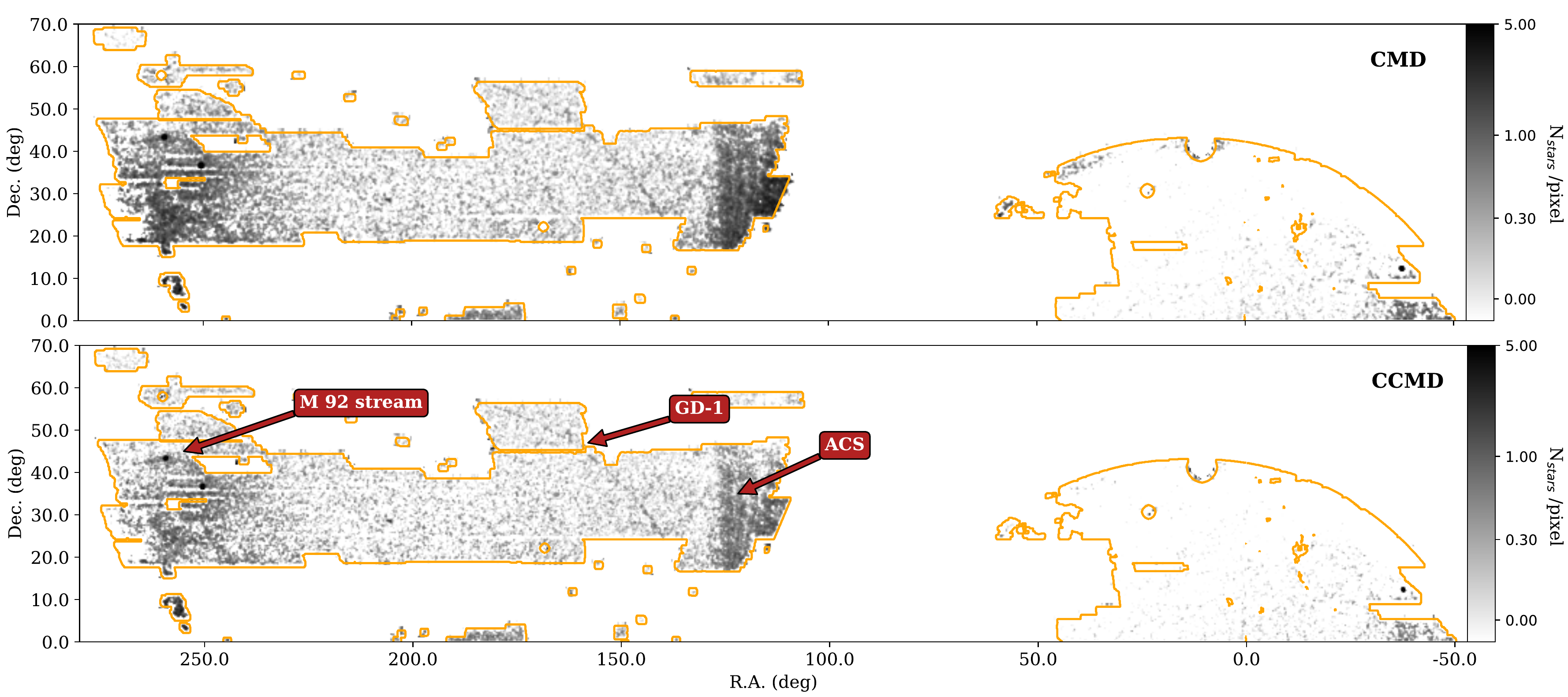}
   \caption{Matched-filter map for a distance of 8 kpc (the distance of M92). The upper panel show the MF conduct using a CMD based filter and the lower panel show the same map using this time a CCMD based filter. The CFIS footprint is indicated by the orange line. The M13 cluster is also visible on this map below M92 on its right side.}
\label{MF_8kpc}
\end{figure*}

The photometric catalogue used in this study is composed of sources observed in the $u$-band of the Canada-France-Imaging-Survey \citep[CFIS][]{ibata_2017a} and in the $g$, $r$ and $i$-bands of the second data release of Pan-STARRS 1 3$\pi$ \citep[][Magnier et al., in prep.]{chambers_2016}.  This catalogue currently covers $\sim 5,200$ deg$^2$ in the northern sky, and is spatially limited by the extent of the current CFIS footprint. The catalogue also contains sources from fields downloaded from the MegaCam archives, hosted by the Canadian Astronomy Data Center (CADC), which were observed  prior to CFIS with the same $u$-band filter ({\sc MP.9302}). The current spatial extent of the catalogue is shown in Figure \ref{MF_8kpc} and is limited by the CFIS footprint indicated in orange.

For the rest of this paper, only stellar-like sources, defined as having $|r_{\mathrm{PSF}}-r_{\mathrm{ap}}| < 0.04$ mag in PS1 are used. It is worth noting that this criterion is more restrictive than the one used by \citet{bernard_2016}, and is a result of the improved reduction process of PS1 DR2 compared to the early Pan-STARRS1 3$\pi$ survey. Our analysis is restricted to objects with individual photometric uncertainties below
0.1 mag in each filter in either $ugr$ or $ugi$. 

The magnitudes of the stars are corrected for foreground reddening by using the extinction values, $E(B-V)$, from \citet{schlegel_1998}. We use the extinction coefficients quoted on the Padova isochrone website\footnote{http://stev.oapd.inaf.it/} for the CFIS\footnote{The PM.9302 correspond to the post-2014 $u$-filter on the Padova website.} and PS1 bands ($ugri$), such that:
\begin{equation}
\begin{array}{ l }
u_0=u-A_v \times 1.50902\\
g_0=g-A_v \times 1.16529\\
r_0=r-A_v \times 0.86813\\
i_0=i-A_v \times 0.67659 , 
\end{array}
\end{equation}
\noindent where $A_v=2.742 \times E(B-V)$ is the absorption coefficient in the $V$-band from \citet{schlafly_2011}.

\subsection{The matched filter}

We first detected the M92 stream in a surface density map obtained by performing a matched-filter (MF) on the CFIS-PS1 catalogue.

The MF \citep{wiener_1949} is a technique used to highlight a specific, known, signal in a noisy dataset. It has been extensively used on large photometric surveys, such as the Sloan Digital Sky Survey (SDSS), PS1 or DES, to discover new thin stellar streams, formed by the disruption of globular clusters (and for a minority of them of dwarf galaxies) around the Milky Way \citep[e.g.][]{rockosi_2002,odenkirchen_2003,grillmair_2006c,balbinot_2011,bernard_2016,shipp_2018}. In doing so, it is assumed that the photometric signal of the stream is similar to the photometric signal of the progenitor globular cluster.  The vast majority of the Galactic globular clusters are well reproduced by old, metal-poor, single stellar populations (SSPs). The MF produces a surface density map which gives higher weight to stars that are more likely to belong to a given SSP than to the field population. The signal is filtered from the background by performing a ratio of the color-magnitude-diagram (CMD; or Hess diagram) of the SSP population to the CMD of field stars. It is possible to probe a range of heliocentric distances by shifting the filter in magnitude-space.

The formalism of the MF used for this work is somewhat similar to the formalism presented in \citet{balbinot_2011} and will be fully described in a future paper (Thomas et al, in prep.). The major innovation is that we use a "color-color-magnitude diagram" (CCMD) instead of a CMD, as visible on Figure \ref{CCMD}. In practice, this means that index $j$ in equations (5, 6, 7) of \citet{balbinot_2011} corresponds to the $j$-th CCMD pixel, instead to the $j$-th CMD pixel. In this work specifically, the MF was carried out in two filter combinations, ($u_0-g_0$, $g_0-r_0$, $r_0$) and ($u_0-g_0$, $g_0-i_0$, $i_0$), which were averaged to produce the final map. The use of a CCMD allows the MF to use the metallicity information encoded in the $u$-band to filter more efficiently the signal of faint stellar streams. The $u$-band photometry is very sensitive to metallicity, due to the high density of metal absorption lines in the near-UV regions \citep{schwarzschild_1955,beers_2005,ivezic_2008,ibata_2017b,thomas_2019a}. Therefore, the $u$-band CCMD reduces the contamination from foreground metal-rich main sequence stars belonging to the Galactic disc that overlaps with the red giant branch population of the more distant metal-poor globular clusters, especially at lower Galactic latitudes, as visible on Figure~\ref{MF_8kpc}. Although, the difference is not drastic, the CCMD map (lower panel) shows that the foreground contamination is sensibly reduce around ($\alpha$,$\delta$)=(250$\degr$,35$\degr$) compare to the CMD map,  carried out in ($g_0-r_0$, $r_0$) and ($g_0-i_0$, $i_0$). Thus, on the CCMD map, structures have a better contrast compared to the foreground. The Anticentre Stream \citep[ACS; ][]{grillmair_2006d} is less pronounced on the CCMD map than on the CMD map. This is because ACS has a metallicity similar to that of the disc ([Fe/H]$=-0.72 \pm 0.26$, \citealt{laporte_2020}), while the MF was conducted for a metallicity of [Fe/H]$\sim$-1.5 (see the next paragraph). Therefore, the fact that the ACS is less pronounced using a CCMD filter shows that it is less affected by the foreground contamination than using a CMD as a filter. Moreover, unlike \citet{bernard_2016}, our formalism takes into account the variation of the CCMD of field stars with Galactic latitude (assuming the Milky Way is axisymmetric). 

As pointed out by \citet{bernard_2016}, synthetic SSPs have many advantages, and are, {\it a fortiori}, better to construct the filter than using an observed globular cluster stellar population, which is subject to contamination from field stars. However, to date, there exist no library of suitable isochrones for the $u$ filter of the CFHT MegaPrime/MegaCam camera, and we have to rely on observed globular clusters in the CFIS footprint to construct the CCMD of the filter. In this paper, we used the globular cluster M13 (NGC 6205) to construct the CCMD of the filter, because this is the closest Galactic globular cluster present in the CFIS footprint, and so has a deeper photometry. Moreover, its photometry is better defined than that of M92. It has a metallicity of [Fe/H]=-1.58 \citep{carretta_2009}, typical for such an object. The same cluster was used by \citet{grillmair_2009} in searches that led to the discovery of the Acheron, Cocytos, Lethe, and Styx stellar streams in SDSS. 

To minimize the impact of differential extinction between different lines of sight, regions with $A_V>0.4$ are masked. This cut remove regions with strong local density variations compared to the rest of the CFIS-PS footprint. Similarly, large known structures (such as the Andromeda, Triangulum and Draco galaxies) are also masked. CFIS is not complete in the center of the M92 cluster due to significant crowding effects in this region. Thus, the inner 4 $r_h$ (i.e. 4.08 arcmin) of the cluster were removed prior to performing the MF. 

\section{Results} \label{sec_results}

\subsection{Analysis of the matched-filter map}  \label{res_MF}

The result of the CCMD MF for a distance of 8 kpc ($m_0-M=14.52$) is presented on the bottom panel of Figure \ref{MF_8kpc}. This image is made with pixels of size $0.1 \degr \times 0.1 \degr$ and smoothed with a $\sigma=0.2 \degr$ Gaussian kernel. The distance of 8 kpc was initially chosen to validate the success of our MF method, because several known structures exist at this distance, including the M13 and M92 globular clusters. On this figure, two known, extended, structures are clearly visible: the GD-1 stream \citep{grillmair_2006a} and the Anticentre Stream \citep{grillmair_2006d,laporte_2020}. In addition to these two structures, a third stream  is visible, emanating from the globular cluster M92 (NGC 6341) and extending over $\sim 17\degr$. A part of this structure ($\sim 5\degr$) was independently reported by \citet{sollima_2020} as this manuscript was being prepared, using Gaia DR2 data \citep{gaiacollaboration_2018}. In that study, only the trailing arm of the stream was detected, whereas both arms can be seen in Figure \ref{MF_8kpc}. This is despite a hole in the CFIS footprint that prevents us from observing the leading arm of the stream (right side arm) beyond $7.5\degr$ from the cluster.

\begin{figure*}
\centering
  \includegraphics[angle=0,viewport= 0 0 1020 410, clip,width=17.0cm]{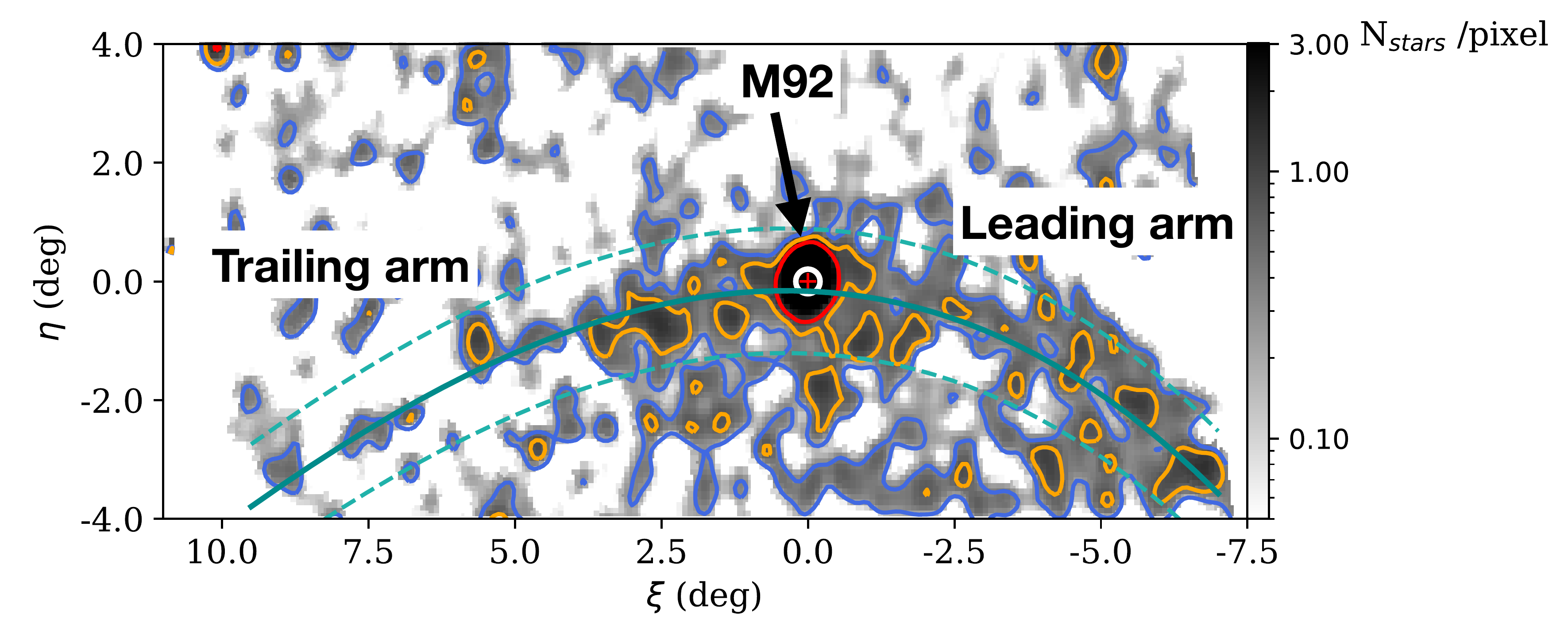}
   \caption{Zoom-in of the CCMD MF signal around M92. The contours represent a 0.3 (blue), 0.7 (orange) and 1.5 (red) stars/pixel. The red cross shows the position of the center of M92 and the white circle show the tidal radius of M92. The cyan line shows the best fit polynomial to the path of the stream, and the dashed lines show the average 3-$\sigma$ width of the stream in the MF ($\sigma=0.35\degr$, after taking into account the Gaussian smoothing). }
\label{fit_M92}
\end{figure*}

\begin{figure}
\centering
  \includegraphics[angle=0, clip,width=7.5cm]{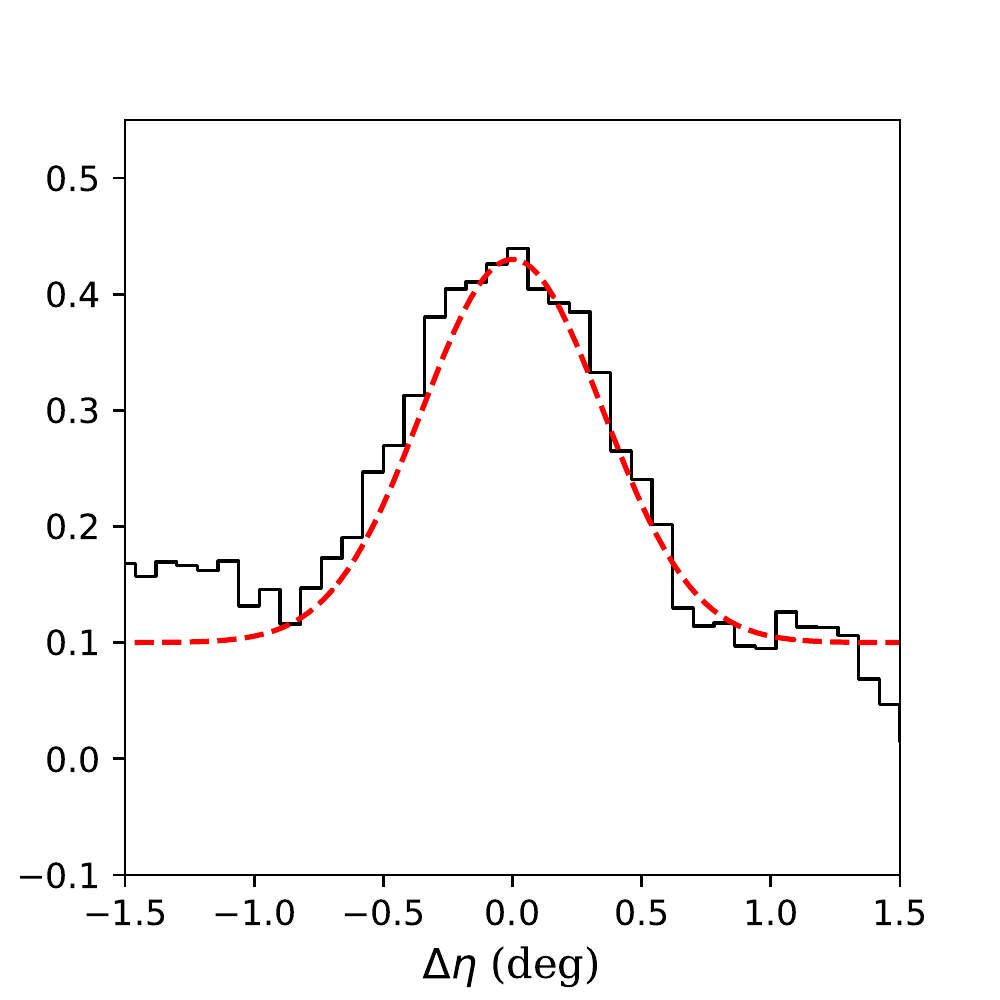}
   \caption{Mean distribution of MF weights perpendicular to the stream in the intervals $-7\degr \le \xi \le -1 \degr$ and $1 \degr \le \xi \le 9.5 \degr$. The dispersion of the fitted Gaussian (red dashed line) is $0.35\degr$. After deconvolution of the smoothing Gaussian, this implies a width of the stream of $\sigma=0.29 \degr$, corresponding to $42\pm{1}$ pc at the distance of M92 \citep[$8.3\pm{0.2}$ kpc, ][]{carney_1992}.}
\label{sigma}
\end{figure}

Figure \ref{fit_M92} presents a zoom-in of Figure~\ref{MF_8kpc} in the region around the M92 globular cluster and its stream. The coordinates of this figure, $(\xi, \eta)$, are in the plan tangential to the celestial sphere at the location of M92. As per convention, $\xi$ increases towards the west and $\eta$ towards the north. In these coordinates, M92 is situating at $(\xi_{M92}, \eta_{M92})=(0\degr,0\degr)$. The presence of a stream on both sides of M92 is very clear. This is despite the fact that on the right side to the cluster (the {\it leading arm}), the contamination from foreground stars (and potentially also from the outskirts of the nearby globular cluster M13), is stronger than on the left side (the {\it trailing arm}) of the cluster. The position of the stream is fitted with a third-order polynomial, only considering pixels with N$_{\rm{stars}}$/pixel$\ge 0.65$, such that:
\begin{equation}
    \eta_{str}(\xi)= -0.134 + 0.041 \ \xi + -0.056 \ \xi^2 + 0.001 \ \xi^3,
\end{equation}
where $\xi$ and $\eta$ are given in degrees.

To quantify the width of the stream, the MF map is co-added in the ranges $-7\degr \le\xi\le-1\degr$ and $1\degr\le\xi\le9.5\degr$ and shown in Figure \ref{sigma}. This region ignores the inner $2\degr$ of the globular cluster so that the main body does not dominate the signal. The red dashed line in Figure~\ref{sigma} shows a Gaussian fit to this distribution and has a dispersion of $\sigma=0.35 \degr$. Taking into account that the MF was smoothed by a Gaussian of $0.2\degr$, this implies a width to the stream of  $\sigma=0.29 \degr$ or $42\pm{1}$ pc at the distance of M92 ($8.3\pm{0.2}$ kpc), slightly larger than the tidal radius of M92 of 30 pc found by \citet{mclaughlin_2005}. A similar width was determined using the unconvolved MF map.

\begin{figure*}
\centering
  \includegraphics[angle=0, clip,width=17.0cm]{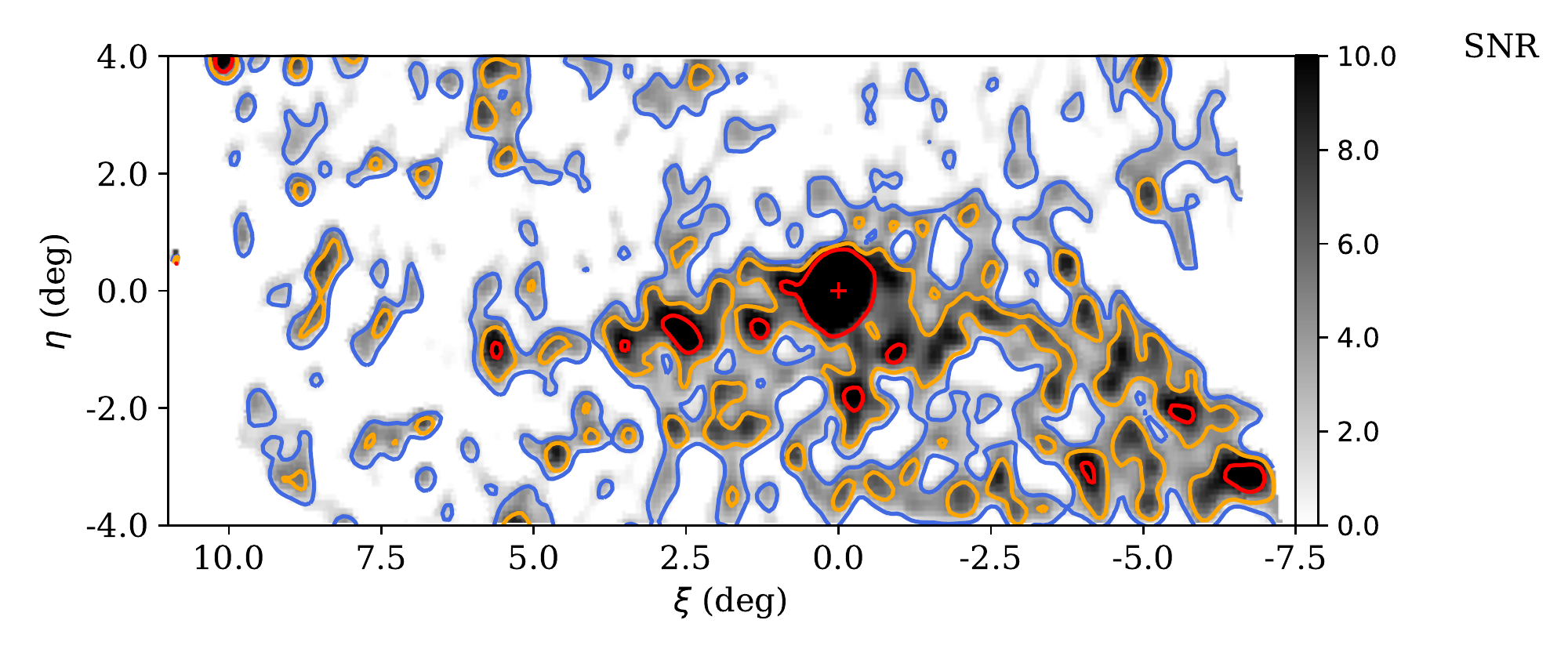}
   \caption{Same as Figure~\ref{fit_M92}, except now expressed as signal-to-noise ratio (SNR) for each pixel, where the background signal is of 0.1 stars/pixel. The contours represent a SNR of 2 (blue), 5 (orange) and 10 (red).}
\label{SNR_M92}
\end{figure*}

In Figure \ref{sigma}, we can see that the number of stars per pixel in the background around the stream is $\sim 0.1$ stars/pixel. The fact it is non-zero is likely due to two factors. The first is that this could correspond to the number of stars in this metallicity range belonging to the "smooth" component of the stellar halo at this distance. Indeed, we note that this is also the average number of stars per pixel in "field" regions at different positions in the MF map at similar Galactic latitudes. However, the second possibility is that there is a residual background/foreground signal in the region around M92 that is due to a non-optimal subtraction of background/foreground stars. This could happen since the MF is constructed using the entire survey region, and not only for the region around M92. In the specific region of M92, there is more contamination from foreground disk stars than at higher Galactic latitudes. If we estimate the background level only very locally, we find that the stream has an average signal to noise of $\simeq 4$. Using a broader area of 4° wide around the fit of the stream to estimate the background level, the average signal to noise is $\simeq 2.3$, due to the presence of the M13 globular cluster, whose distance of $7.1\pm{0.1}$ kpc \citep{deras_2019} is close of the $8.3 \pm{0.2}$ kpc of M92, and so is visible on the MF map due to the intrinsic scatter of its CCMD.

The SNR for each pixel is shown in Figure \ref{SNR_M92}. The stream is clearly visible stretching from each side of the cluster, despite the leading arm (right side) being less well defined than the trailing arm due to an increase of the contamination, as mentioned above. As we will see later (Section \ref{sec_simu}), it is actually possible that the stream becomes wider beyond $\sim 4\degr$.

Following \citet{ibata_2017}, we estimate the mass of the stream by comparing the MF counts in the stream to those within the tidal radius of the globular cluster ($r_t$). This is not straightforward, because the inner 4 $r_h$ of the cluster is affected by crowding. However, under the reasonable assumption that M92 follows a King profile described by the parameters reported by \citet{mclaughlin_2005}, $12.5\%$ of the mass of the cluster is between $4 r_h$ and $r_t$. Additionally, the CFIS data in the inner South-West half of the M92 cluster suffers from poor data processing and calibration, and so we do not use it to estimate the mass of the stream. Instead, we use only the North-East half of the cluster to estimate the mass. By correcting for the missing $87.5\%$ of the stars, we find the ratio in stellar mass between the stream (within its 3-$\sigma$ width along the polynomial fit) and the main body of the cluster to be $0.10\pm{0.02}$. From the parameters listed in Table \ref{param_M92}, we estimate the mass of the cluster to be of $[3.17\pm{0.26}] \times 10^{5}$ M$_\odot$, which leads to a mass of the stream of $[3.17 \pm{0.89}] \times 10^{4}$ M$_\odot$. Note that we expect that the formal uncertainty quoted above is likely an underestimate, and that this mass corresponds only to that part of the stream that we can clearly detect. This general point is especially relevant for M92, since the proper motion for M92 suggests that its orbit takes it through the bulge of the Milky Way, and could be perturbed by the Galactic Bar\footnote{\citet{baumgardt_2019} estimate the pericenter to be at $\sim 2$ kpc, although the exact value depends on the choice of the potential}. This means that it is possible that some stars from M92 are on chaotic orbits and are not present along the thin stream that we detect \citep{pearson_2015,price-whelan_2016,hattori_2016,bonaca_2020}.

 \begin{table}
 \centering
  \caption{Properties of the globular cluster M 92. The sources are : $1 =$ \citet{goldsbury_2010}, $2 =$ \citet{carney_1992}, $3 =$ \citet{baumgardt_2019}, $4 =$ \citet{carretta_2009}, $5 =$ \citet{harris_1996,harris_2010}, $6 =$ \citet{mclaughlin_2005}.}
  \label{param_M92}
  \begin{tabular}{@{}lcc@{}}
  \hline
   Parameter & Value & Source  \\
    \hline
   RA & $17^h17^m07.39^s$ & 1 \\
   Dec & $+43\degr 08'09.4"$ & 1 \\
   Distance & $8.3 \pm{0.2}$ kpc & 2 \\
   V$_{rad}$ & $-120.48 \pm 0.27$ km.s$^{-1}$ & 3\\
   $\mu_\alpha$ & $-4.93\pm{0.2}$ mas.yr$^{-1}$& 3 \\
   $\mu_\delta$ & $-0.57\pm{0.2}$ mas.yr$^{-1}$& 3 \\
   $[$Fe/H$]$ & $-2.35\pm{0.05}$ & 4 \\
   M$_{\mathrm{v}}$ & $-8.21$ & 5 \\
   $\gamma_{\mathrm{v}}$ & $1.93  \pm{0.16}$ M$_\sun$.L$_\sun^{-1}$ & 6 \\
   $r_{c}$ & 0.26 arcmin  & 6 \\
   $r_t$ & 12.44 arcmin & 6 \\
   $r_h$ & 1.02 arcmin & 6 \\
   Mass & $[3.17\pm{0.26}] \times 10^5$ M$_\odot$ & This work \\
\hline
\end{tabular}
\end{table}

\subsection{Confirmation using other tracers} \label{sec_other}

To further confirm the presence of the stream emanating from M92, we compare the position of the stream detected on the MF map with that of stars from other catalogues that are bright enough to have proper motion measurements from Gaia.

\begin{figure*}
\centering
\includegraphics[angle=0,viewport= 0 45 565 185, clip,width=17.0cm]{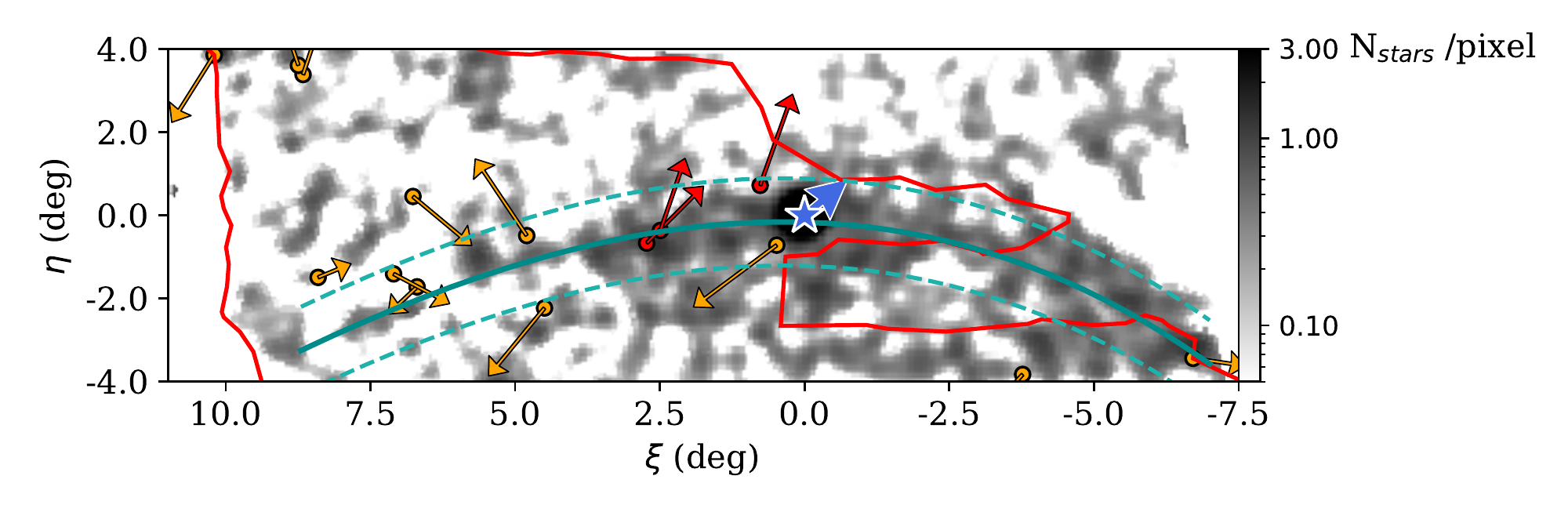}
\includegraphics[angle=0,viewport= 0 45 565 185, clip,width=17.0cm]{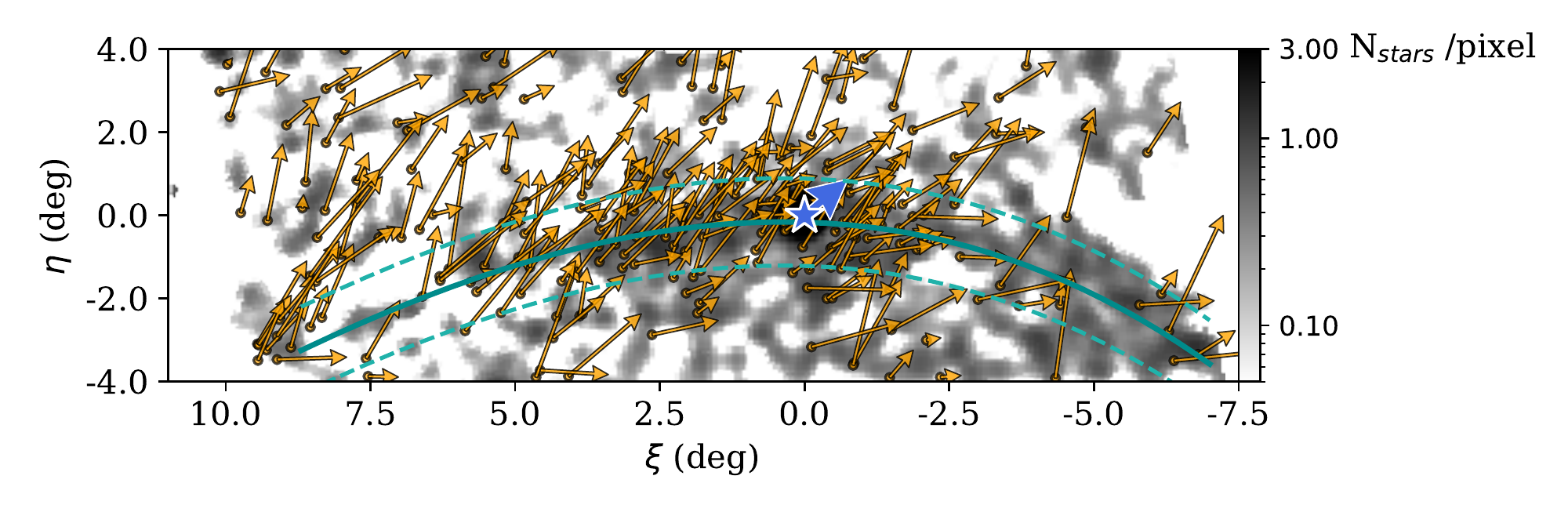}
\includegraphics[angle=0,viewport= 0 0 565 185, clip,width=17.0cm]{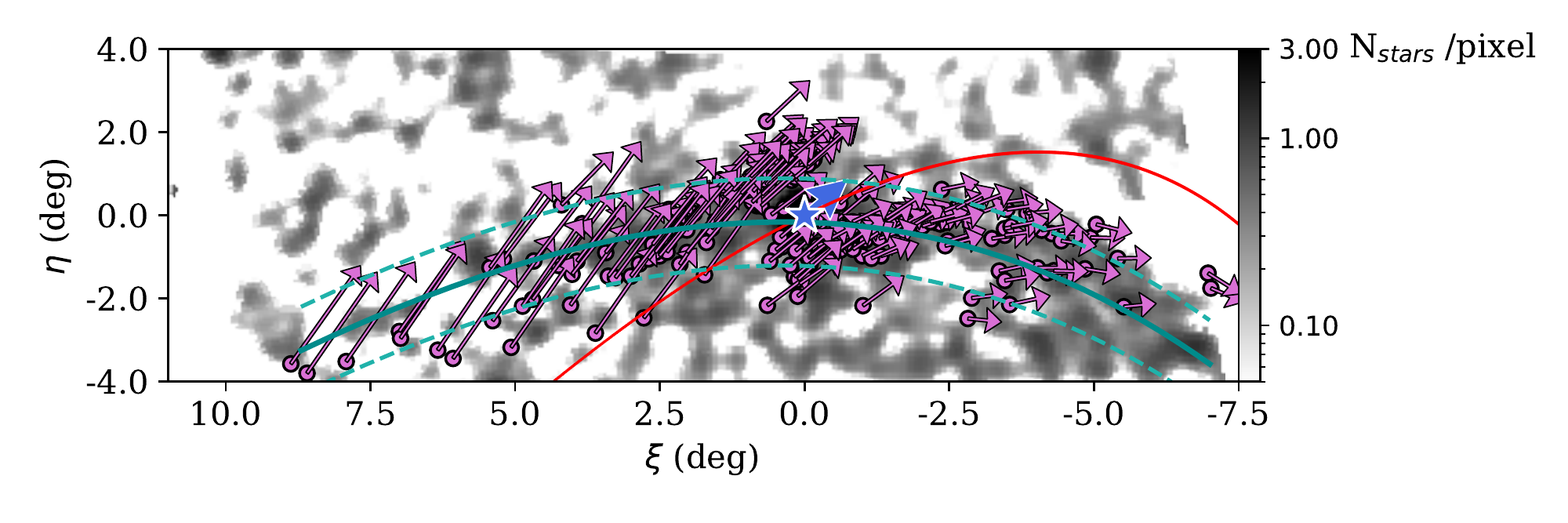}
   \caption{{\it Upper panel:} The dots represent the position of BHBs around M92 that satisfy the criteria described on Section \ref{BHB}, and the arrows show their proper motion (corrected for the Solar reflex motion). The red dots and arrows highlight the 3 BHBs that are likely member of the M92 stream. The red line shows the spatial limits of the catalogue of the BHB catalogue of \citet{thomas_2018a}.
   {\it Middle panel:} MSs and RGBs that sastisfy the criteria listed on Section \ref{sec_DG} (in orange). {\it Lower panel:} Particles spray from the model describe in section \ref{sec_mod} (in pink). The red line show the orbit of the cluster in the potential used by this model. For each of these panels, the background grayscale image correspond to the MF map of Figure \ref{fit_M92}, the blue star show the position of the globular cluster and the blue arrow its mean proper motion. The cyan line shows the polynomial best fit to the position of the stream, and the dashed lines show the typical 3-$\sigma$ width of the stream in the MF.}
\label{M92_cats}
\end{figure*}

\subsubsection{Blue horizontal branch stars} \label{BHB}

We first compare the MF map to the Blue Horizontal Branch (BHB) catalogue of \citet{thomas_2018a}, whose distances have been measured with a relative precision of $\simeq 10\%$ using the relation between their absolute magnitude and their $(g-r)_0$ color provided by \citet{deason_2011}. The upper panel of Figure \ref{M92_cats} shows BHBs around M92, in the range $7.3\le d_{helio} \le 9.3$ kpc and with a proper motion of maximum twice that of M92 ($|\mu|<2 |\mu_{M92}|$). This last criterion is broad enough to take into account that the individual uncertainties on the proper motion are comparable to as the measurements themself for stars at the distance of M92. It have to be noted that the BHB catalogue of \citet{thomas_2018a} was used with a previous data release of CFIS that was not as extended as the present one, and its footprint in the M92 region is shown by the red line. For clarity, the BHBs inside the cluster are not shown. Arrows show the proper motion of the stars and the blue arrow shows the mean proper motion of M92 found by \citet{baumgardt_2019}. This is listed with the other parameters of M92 in Table \ref{param_M92}. Proper motions are corrected for the Solar reflex motion, assuming that the Sun is at a distance of $8.129$ kpc from the Galactic center \citep{gravitycollaboration_2018a}. The circular velocity is assumed to be $229.0$ km.s$^{-1}$ \citep{eilers_2019}, and we use the adopted Solar peculiar motion from \citet{schonrich_2010}, namely ($U_\odot$, $V_\odot$, $W_\odot$) = (11.1, 12.24, 7.25) km.s$^{-1}$ in Local Standard of Rest coordinates. 

It is interesting to note in Figure~\ref{M92_cats} that the mean proper motion of the cluster is not aligned with the stream, as is common for most globular clusters streams \citep[e.g.][]{price-whelan_2018,malhan_2018,ibata_2020}. This is because M92 is just before its apocenter (as indicated by the path of the red line in the lower panel of Figure \ref{M92_cats}). Indeed, the stars on the leading arm have a lower potential energy that the stars remaining in the cluster, and thus have a slightly closer apocenter than them. The inverse is true for stars in the trailing arm. Therefore, {\it at this specific location, the stream is not aligned with the orbit of the cluster}, with an angle between the orbit of the cluster and the fitted position of the stream (i.e. the angle between the cyan and red lines on the lower panel of Figure~\ref{M92_cats}) of $\theta_{M92}=40 \degr$ (at the position of the cluster). Thus, most of the non-aligned velocity are caused by the precession of the orbital plane of M92.

For each BHB, we compute the angle ($\theta$) between their apparent motion and the fitted position of the stream at their position. We can then define likely members of the stream  as those stars that go in the same general direction of the cluster ($|\theta-\theta_{M92}| < 45 \degr$) and are within 3-$\sigma$ of the width of the stream. Three BHBs match these criteria and are highlighted in red in the upper panel of Figure \ref{M92_cats}. All of them are located in the trailing arm, two of them are very close to the fitted position of the stream and the third one is close to the possible location of Lagrange point L2. Despite being a very sparse tracer population, BHBs have the advantage among other stellar tracers to have precise distance measurements (10\% precision), and so can be used as reliable tracers to confirm the existence of the stream.

\subsubsection{Main sequence and red giant branch stars} \label{sec_DG}

To supplement the BHB catalog, we also consider main sequence (MS) and red giant branch (RGB) stars from the catalogue of \citet{thomas_2019a}. The metallicities and distances of the stars from this catalogue have been derived photometrically. Stars from this catalogue that satisfy the following criteria are shown in the middle panel of Figure \ref{M92_cats} :
\begin{itemize}
    \item $-2.5 \le [$Fe/H$] \le -2.0$
    \item $7.3 \le d_{helio} \le 9.3$ kpc
    \item $|\mu| < 2 |\mu_{M92}|$
    \item $|\theta-\theta_{M92}| < 45 \degr$
    \item $\varpi-2 \delta \varpi \le 1.0/7.3'$
    \item $\sqrt{\delta \mu_{\alpha}^{*2}+ \delta \mu_{\delta}^2} < 4.0$ mas.yr$^{-1}$.
\end{itemize}
$\varpi$ is the Gaia parallax corrected from the zero point offset of 0.029 mas.yr$^{-1}$ \citep{lindegren_2018}, $\delta \varpi$ is the uncertainty on the parallax, $\mu$ is the proper motion\footnote{corrected from the Solar reflex motion} of the stars, and $\mu_{M92}$ is the global proper motion of the M92 cluster. 

The first of the above criteria remove the majority of metal-rich foreground Galactic disk stars and the second and third criteria are the same as used for the BHBs. The fourth criterion retains only those stars going in general the same direction as the cluster. The last two criteria remove fewer than 2\% of the stars by excluding the few nearby stars with good Gaia parallaxes that clearly have an incorrect photometric distance, as well as those with very poorly determined proper motions. 

The middle panel of Figure \ref{M92_cats} clearly shows that the large majority of stars that satisfy these criteria are located along the stream, with a density $3 - 4$ times higher that of the field. Most of these stars are located in the trailing arm. However, the leading arm is well populated out to $\sim 2.5$ degrees from the cluster.  

The lower number of kinematically-selected stars in the leading arm compared to the trailing arm could be a consequence of a wrong fit to the position of the leading arm, since the contamination is more important in this region than in the trailing arm, leading to a miscalculation of the angle $\theta$. Another explanation could be inherent to the CFIS photometry used by \citet{thomas_2019a} to make this catalogue of stars, since the CFIS $u$-band photometry has a more uncertain zero point calibration in this region of sky. An error on the zero point calibration could lead to wrong estimates of the photometric metallicities and of the distances derived by \citet{thomas_2019a}. In this eventuality, the MF will be less affected due to the use of a relatively wide filter to define the signal (which therefore does not require very precise photometry). In short, we urge caution in drawing robust conclusions from the low number of kinematically selected stars in the leading arm at this stage.

\section{Dynamical modelling of the stream} \label{sec_dynmod}

We now undertake dynamical modelling of M92 and its stream, to attempt to understand its dynamical age and orbital properties. The presence of a remnant cluster greatly facilitates the simulation of the stream by reducing the number of free parameters, especially concerning the orbit of the progenitor, in contrast to  "progenitor-free" streams like GD-1 \citep{grillmair_2006a}. We now describe two different models of the stream, the first created by spraying particles at the Lagrange points \citep{varghese_2011}, and the second using a full N-body simulation. 

 \begin{figure*}
\centering
\includegraphics[angle=0,viewport= 12 16 565 190,clip,width=17.0cm]{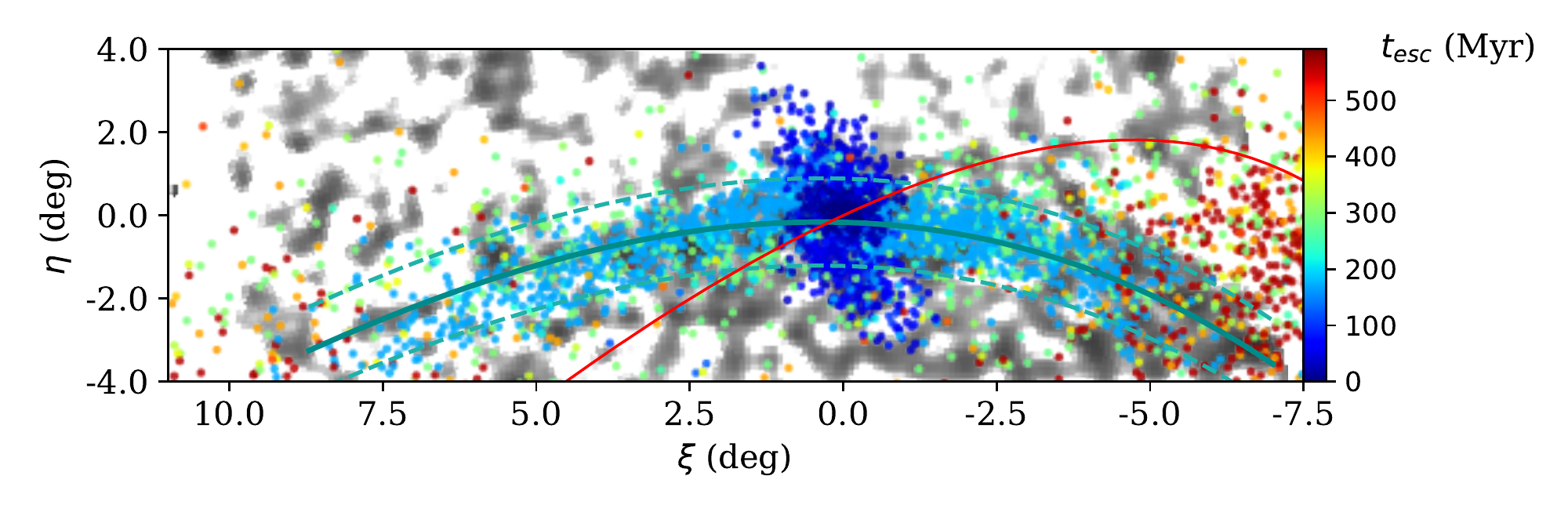}
   \caption{Projection of the particles of the simulation of M92 described in Section \ref{sec_simu} overlaid on the MF. The particles are color-coded by look-back time relative to when they escaped the progenitor. The red line shows the orbital path of the M92 cluster.}
\label{M92_simu}
\end{figure*}

\subsection{Spraying particles} \label{sec_mod}

 Here, we use the \textsc{gala} package \citep{price-whelan_2017} to model the stream by spraying particle at the Lagrange points at every time step ($dt = 5$ Myr), using the distribution function developed by \citet{fardal_2015}. 
 
 The Milky Way potential is modelled by a Miyamoto-Nagai disk with a mass of $5.56 \times 10^{10}$ M$_\odot$, a scale length of $a=3.5$ kpc and a scale height of $b=0.28$ kpc. It also includes a NFW halo \citep{navarro_1997} with a Virial mass of $0.84 \times 10^{12}$ M$_\odot$ and a scale length of $r_s=17.19$ kpc. This produces a circular velocity at the Solar radius of 229.3 km.s$^{-1}$, consistent with the value found by \citet{eilers_2019} that we previously used to correct the PM of the Solar reflex motion. Our model uses the present-day position and velocity of the globular cluster, listed in Table \ref{param_M92}. Although it does not affect significantly the dynamics of the stream, we include the self-gravity of the cluster by adding the potential of a \citet{plummer_1911} sphere of mass $3.17 \times 10^{5}$ M$_\odot$ with a scale radius of $2.4$ pc.
 
 The position and proper motion of the particles generated by this model are compared to the MF in the lower panel of Figure \ref{M92_cats}. The large majority of these particles have been sprayed very recently, in the last $300-350$ Myr. All of them were sprayed less than $\simeq 500$ Myr ago. Since the M92 cluster has an orbital period of $\simeq 130$ Myr, this implies that the stream has been formed over the last $4 - 5$ orbits, with most of the stars in the stream having escaped during the last orbit. Using these timescales and the mass of the stream found in Section \ref{res_MF}, it is possible to conclude that the cluster lost on average $\simeq 6.3 \times 10^4$ M$_\odot.$Gyr$^{-1}$. If this rate is constant, M92 will be fully disrupted in the next 5 Gyr. However, due to the loss of mass, its tidal radius will become smaller, and so it is very likely that the cluster will be completely disrupted in the next $1 - 2$ Gyr \citep[see][]{meiron_2020}.

We also note that, with this model, we can validate the selection criteria used in Sections \ref{BHB} and \ref{sec_DG}, since most of the particles sprayed over the last 500 Myr appear to respect these criteria. The particles that do not respect these criteria have been ejected from the stream due to repeated pericentric passages of the cluster close to the Galactic center.

 \begin{figure}
\centering
\includegraphics[angle=0,width=8cm]{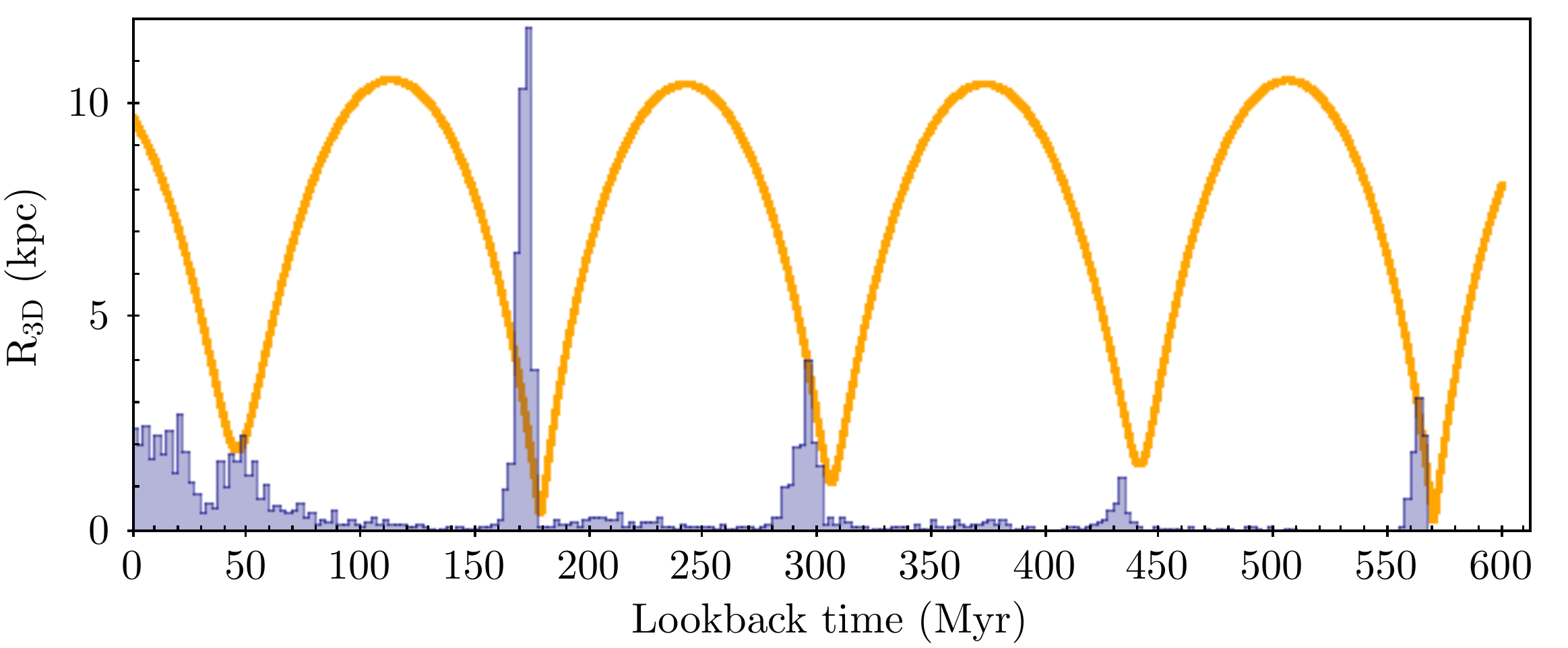}
   \caption{The three dimensional separation of the progenitor as a function of lookback time for the N-body simulation is shown in orange. Also overlaid as a histogram is the relative count of stars escaping the cluster as a function of the lookback time.}
\label{orbite}
\end{figure}

\subsection{N-body simulation} \label{sec_simu}

We have also performed a full non-collisional N-body simulation of the disruption of M92, using the GyrfalcON integrator \citep{dehnen_2000,dehnen_2002} that is part of the {\sc Nemo} package \citep{teuben_1995}. The choice to use a non-collisional instead of a fully collisional code was made to reduce the computational time, but is also justified by the fact that \citet{meiron_2020} recently showed that internal two-body encounters do not play a major role in the dissolution of a massive globular cluster like M92. The adopted Galactic potential for this simulation is the same as the one used by \citet{ibata_2020} to simulated the GD-1 stream. This potential is composed of a bulge, thin disk, thick disk and interstellar medium of model 1 of \citet{dehnen_1998}. The dark matter halo is similar to the halo found by \citet{cautun_2020}, constructed using a \citet{navarro_1997} profile, with a virial radius of 206 kpc, a concentration of $c = 12$, and with an oblateness of $q = 0.82$ (\citealt{malhan_2019}). This Galactic potential model has a circular velocity at the Solar
radius of 229 km.s$^{-1}$, consistent with the value found by \citet{eilers_2019} that we used earlier.

To find its initial position for the simulation, the M92 globular cluster was integrated backward from its current position (listed in Table \ref{param_M92}) for 600 Myr. We then  integrate it forward using a \citet{king_1966} model with a mass M$_{gc}$ = $3.8 \times 10^5$ M$_\odot$, a core radius of $r_c= 1.5$ pc, and a ratio between the central potential and the velocity dispersion of $W_0 =
7.5$. These parameters were set to produce a stream with a mass consistent with $3.1 \times 10^4$ M$_\odot$, as found in Section \ref{res_MF}, while also having a remnant cluster with similar properties to the current M92. The cluster is modelled with $32,000$ equal-mass particles and the adopted smoothing scale length in GyrfalcON is $0.5$ pc (due to the size of the cluster).

The spatial distribution of particles at the end of the simulation, projected on the $(\xi, \eta)$ plane and color-coded by the time when they escaped the progenitor, are shown in Figure \ref{M92_simu}. As was the case in the spraying-particle model, the bulk of the stars in the stream were ejected in the last 300 Myr. Indeed, 50\% of them were ejected just after the penultimate pericentric passage of the cluster at the pericenter, shown in Figure \ref{orbite}, which also shows the change in Galactocentric radius as a function of time over the orbit. We also note that all the particles along the detected part of the stream were ejected within the last 600 Myr, even though we did initially run simulations over a longer period of time. However, none of these produced particles have a position consistent with the observed stream. This confirms our conclusion from the particle spraying analysis, which is that the stream is a relatively recent creation, with an age of $\sim 500$ Myr. 

The initial mass of the progenitor that we used was slightly more massive than the current total mass of the system (stream + cluster) that we previously derived. This accounts for the fact that most of the stars that escaped at the first pericenter (at a lookback time of 570 Myr) are not distributed along the path of the stream that we detected. Rather, most of these stars are fanned over a wider area, similar to the ``fan`` structure recently observed along the Palomar 5 stream \citep{bonaca_2020}. The stars composing this structure are on a slightly different orbit than M92's. If such a structure is indeed present along the M92 stream, it will be a very low surface brightness structure that would be very difficult to detect, especially taking into account that this region is close to the Galactic disk. We tentatively note that the phase-space dispersion linked to a possible  ``fanning'' of the stream could also partially explain why the region around the leading arm is more spread out than in the trailing arm (in addition to the stronger contamination in this region that we previously discussed).

\section{Discussion} \label{sec_discussion}

 It is very interesting to find that the M92 stream has a dynamical age of $\sim 500$ Myr, while the M92 cluster hosts a stellar population aged of $11\pm{1.5}$ Gyr \citep{dicecco_2010}. It is possible that the M92 stream, as currently detected, is the tip of the iceberg of a more diffuse structure formed from stars that escaped the cluster at earlier time. Although such a diffuse structure would have a very low-surface brightness and would likely be hard to detect. However, it is also possible that the difference between the dynamical age of the stream and the age of the stellar population in its progenitor is directly linked to the origin of M92.
 
 At this stage, several interesting possibilities emerge:
\begin{enumerate}
    \item  Since M92 has recently passed close to the Galactic center, including possibly interacting with the Galactic bar, it is possible that M92 was not originally on such a disruptive orbit and has only recently been thrown on its current orbit;
    
    \item M92 could have been brought into the Galaxy by a dwarf galaxy, which will have suffered from orbital decay due to the dynamical friction with the Galactic dark matter halo \cite[e.g.][]{chandrasekhar_1943,cora_1997}. This host is now either completely destroyed or on a completely different orbit \citep[see][]{malhan_2019a,malhan_2020};
    
    \item An alternative to the previous point is that M92 is the remnant nucleus of the progenitor galaxy, rather than being one of its globular cluster \citep[e.g.][]{searle_1978,freeman_1993,boker_2008}. Based on result from the Next Generation Virgo Cluster Survey \citep[NGVS;][]{ferrarese_2012}, if M92 is the remnant nucleus of a dwarf galaxy, this galaxy would have a metallicity of $[$Fe/H$]\sim -2.1$ \citep{spengler_2017}, a mass of $ M= 10^{7\pm{1}}$ M$_\odot$ and an effective radius between 250 and 900 pc \citep{sanchez-janssen_2019a}.
\end{enumerate}

At this date, we did not find any traces of a disrupted dwarf galaxy close M92. However, in the future, we plane to explore the different space parameters, especially the metallicity and dynamical space, using jointly the CFIS, PS1, Pristine \citep{starkenburg_2017} surveys and the incoming Gaia early data release 3. In parallel, we plane to make a more detailed model of the cluster and of its environment, especially by accounting for the presence of the Galactic bar in the Galactic potential.

\section{Summary} \label{sec_conclusion}

We report on the discovery of a stellar stream emanating from the globular cluster M92 (NGC 6341) using photometry from CFIS and the PS1 survey. Part of this stream was independently detected by \citet{sollima_2020} using Gaia DR2 data during the preparation of this manuscript. Our detection of the M92 stream was made possible by using the metallicity information contained in CFIS $u$-band to improve the match-filtering technique, and by taking into account the spatial variation of the Galactic foreground population. 

The detected stream has a projected length of $\simeq 17 \degr$  (or $ \simeq 2.5$ kpc at the distance of M92) and a width of 0.29$\degr$ (42 pc). We find that the detected portion of the stream has a mass of $[3.17 \pm{0.89}] \times 10^{4}$ M$_\odot$, about $10 \%$  the mass of the current main body of M92. Moreover, we confirm the existence of the M92 stream kinematically with main sequence, red giant and blue horizontal branch stars, all of which have Gaia proper motion measurements.

We also present dynamical modeling of the stream using two different methods, by regularly spraying particles at the Lagrange points and with a realistic, non-collisional, N-body simulation. Both models show that the stream seems to have been formed very recently, during the last $\sim 500$ Myr, with most of the it being younger than 370 Myr. This observation is very interesting since the M92 cluster is one of the oldest and most metal-poor globular cluster around the Milky Way \citep[e.g.][]{harris_1996,harris_2010}, forcing us to question the origin of this cluster. 

At this stage, several interesting possibilities emerge:
\begin{enumerate}
    \item The M92 stream as currently detected could be the tip of the iceberg of a more diffuse structure;
    \item The orbit of M92 may have change recently, possibly due to an interacting with the Galactic bar; 
    \item M92 may previously have been brought into the Galaxy by a dwarf galaxy, which is either now completely destroyed or on a completely different orbit.
    \item M92 is the remnant nucleus of a dwarf galaxy.
\end{enumerate}

Investigating these interesting possibilities will require a more detailed model of the cluster, likely taking into account its collisional nature and the presence of the Galactic bar in the Milky Way potential. Certainly, this stream appears to be a potentially very valuable beacon to probe the inner  three dimensional structure of the Galactic potential.

\section*{Acknowledgments}

We thanks Todd Burdullis and the all QSO team for the care and dedication given to planning and observing this survey, providing us these fantastic data. We also thanks Eugene Magnier for useful insights on the PS1 photometry.

NFM gratefully acknowledge support from the French National Research Agency (ANR) funded project ``Pristine`` (ANR-18-CE31-0017) along with funding from CNRS/INSU through the Programme National Galaxies et Cosmologie and through the CNRS grant PICS07708. 

KM acknowledges support from the $\rm{Vetenskapsr\mathring{a}de}$t (Swedish Research Council) through contract No. 638-2013-8993 and the Oskar Klein Centre for Cosmoparticle Physics.

ES gratefully acknowledges funding by the Emmy Noether program from the Deutsche Forschungsgemeinschaft (DFG).

This work is based on data obtained as part of the Canada-France Imaging Survey (CFIS), a CFHT large program of the National Research Council of Canada and the French Centre National de la Recherche Scientifique. Based on observations obtained with MegaPrime/MegaCam, a joint project of CFHT and CEA Saclay, at the Canada-France-Hawaii Telescope (CFHT) which is operated by the National Research Council (NRC) of Canada, the Institut National des Science de l'Univers (INSU) of the Centre National de la Recherche Scientifique (CNRS) of France, and the University of Hawaii, and on data from the European Space Agency (ESA) mission {\it Gaia} (\url{https://www.cosmos.esa.int/gaia}), processed by the {\it Gaia} Data Processing and Analysis Consortium (DPAC, \url{https://www.cosmos.esa.int/web/gaia/dpac/consortium}). Funding for the DPAC has been provided by national institutions, in particular the institutions participating in the {\it Gaia} Multilateral Agreement.

We also used the Pan-STARRS1 Surveys (PS1), that have been made possible through contributions of the Institute for Astronomy, the University of Hawaii, the Pan-STARRS Project Office, the Max-Planck Society and its participating institutes, the Max Planck Institute for Astronomy, Heidelberg and the Max Planck Institute for Extraterrestrial Physics, Garching, The Johns Hopkins University, Durham University, the University of Edinburgh, Queen's University Belfast, the Harvard-Smithsonian Center for Astrophysics, the Las Cumbres Observatory Global Telescope Network Incorporated, the National Central University of Taiwan, the Space Telescope Science Institute, the National Aeronautics and Space Administration under Grant No. NNX08AR22G issued through the Planetary Science Division of the NASA Science Mission Directorate, the National Science Foundation under Grant No. AST-1238877, the University of Maryland, and Eotvos Lorand University (ELTE).

\bibliography{./biblio}

\end{document}